\documentclass[conference]{IEEEtran}
\IEEEoverridecommandlockouts
\usepackage{cite}
\usepackage{amsmath,amssymb,amsfonts}
\usepackage{graphicx}
\usepackage{subfigure}
\usepackage{textcomp}
\usepackage{xcolor}
\usepackage{algorithm,algorithmic}
\usepackage{bm}
\usepackage{booktabs}
\usepackage{multirow}
\usepackage{geometry}
\newgeometry{
  top=72pt,
  bottom = 54pt,
  left = 54pt,
  right = 54pt
}
\def\BibTeX{{\rm B\kern-.05em{\sc i\kern-.025em b}\kern-.08em
    T\kern-.1667em\lower.7ex\hbox{E}\kern-.125emX}}
\begin{document}
%\title{Learning to Collaboratively Operate Shared Energy Storage: A Multi-Agent Approach }
\title{Learning a Multi-Agent Controller for Shared Energy Storage System}

\newcommand\ChangeRT[1]{\noalign{\hrule height #1}}

\author{\IEEEauthorblockN{Ruohong Liu and Yize Chen}
\IEEEauthorblockA{\textit{Artificial Intelligence Thrust, Information Hub} \\
\textit{The Hong Kong University of Science and Technology (Guangzhou)}\\
%Guangzhou, China \\
rliu519@connect.hkust-gz.edu.cn, yizechen@ust.hk}
}

\maketitle

\begin{abstract}
    Deployment of shared energy storage systems (SESS) allows users to use the stored energy to meet their own energy demands while saving energy costs without installing private energy storage equipment. In this paper, we consider a group of building users in the community with SESS, and each user can schedule power injection from the grid as well as SESS according to their demand and real-time electricity price to minimize energy cost and meet energy demand simultaneously. SESS is encouraged to charge when the price is low, thus providing as much energy as possible for users while achieving energy savings. However, due to the complex dynamics of buildings and real-time external signals, it is a challenging task to find high-performance power dispatch decisions in real-time. By designing a multi-agent reinforcement learning framework with state-aware reward functions, SESS and users can realize power scheduling to meet the users' energy demand and SESS's charging/discharging balance without additional communication, so as to achieve energy optimization.  Compared with the baseline approach without the participation of the SESS, the energy cost is saved by around $2.37\%$ to $21.58\%$.
\end{abstract}

\vspace{-0.5em}
\section{Introduction}
Energy storage is gaining more attention since it enables higher penetration of renewables, achieving energy arbitrage and enhancing the power systems resilience \cite{divya2009battery, dai2021utilization}. However, the high installation and maintenance costs of energy storage systems hinder their application \cite{bhusal2019optimum}. In contrast, installing a shared energy storage system (SESS) for the community is a more economical and feasible solution, because users can maximize the utilization of energy storage without installing an individual energy storage system. The California Public Utilities Commission (CPUC) has broadly defined community storage as storage connected at the distribution feeder level~\cite{CPUC2015}, associated with a cluster of customer load. The services of these types of systems could provide capacity for excess generation from distributed energy resources (DERs) and backup power during outages. Previous works have demonstrated that shared energy storage could achieve electricity cost saving with higher utilization compared to individual energy storage \cite{walker2021analysis}.

Most of these existing studies on SESS assume that users' demand patterns as well as energy prices are known ahead of time or can be predicted perfectly~\cite{yang2021optimal}, thus the system operator only needs to optimize the charging of SESS and power dispatch to each user \cite{zhu2021distributed}. However, due to time-varying external signals such as electricity prices and complex user demand behaviors such as heating, ventilating and air conditioning (HVAC) systems,  these assumptions are not realistic in practice. Therefore, we propose to consider the setting where multiple users learn to cooperatively operate a SESS to simultaneously reduce energy cost and satisfy energy demand in a distributed manner. %Such an approach is made practical as all users could make use of the shiftable demands and implement energy storage arbitrage based on price signals.

Simultaneously controlling the charging and discharging of the SESS and the energy demand of the users is a complex problem. Due to the limited capacity of the SESS, charging and discharging operations will affect the subsequent decisions, which makes the problem a time-coupling problem. It is hard to model such complex dynamical systems using classical dynamic programming techniques\cite{zhu2021distributed}. Furthermore, centralized control or reinforcement learning schemes require system operators to know their full state, which places high demands on communication and raises privacy issues when collecting the state information from individuals~\cite{odonkor2018control}. In our proposed distributed control approach, the energy storage system only needs to respond to price signals without predicting the energy demand of the building, while building user agents decide discharging from SESS and power injection from the grid. With multi-agent reinforcement learning (MARL) setup  ~\cite{hu1998multiagent, lowe2017multi}, the multi-objective optimization problem can be simplified as maximizing each agent's reward.
%\textcolor{red}{Describe two motivation explicitly. Why it is important to control building+battery together? Why solving this problem is hard (complicated dynamics, agent interactions,...)?}

In this paper, we consider the scenario with the collocation of energy storage and residential homes. We are interested in integrating energy storage while shifting the portion of their electricity demand load in response to time-varying electricity price, i.e., demand response. We design a collaborative learning approach based on MARL. States and actions information of all agents are provided to each agent during the training process to help each policy neural network learn the policy, while only the local state is utilized during distributed execution. We observe reward shaping are important for MARL training, where negative rewards are proposed for temperature derivation and energy cost for building agents; meanwhile, rewards are designed to encourage the energy storage can utilize electricity price fluctuations. We test our proposed method with real-world temperature and price based on a group of realistic building models, and the performance shows that around $2.37\%$ to $21.58\%$ of energy cost is saved compared to the baseline setup without the participation of the SESS.

\vspace{-0.5em}
\section{Problem Setup}
\begin{figure}[htbp]
\vspace{-5pt}
\centerline{\includegraphics[width=0.5\textwidth]{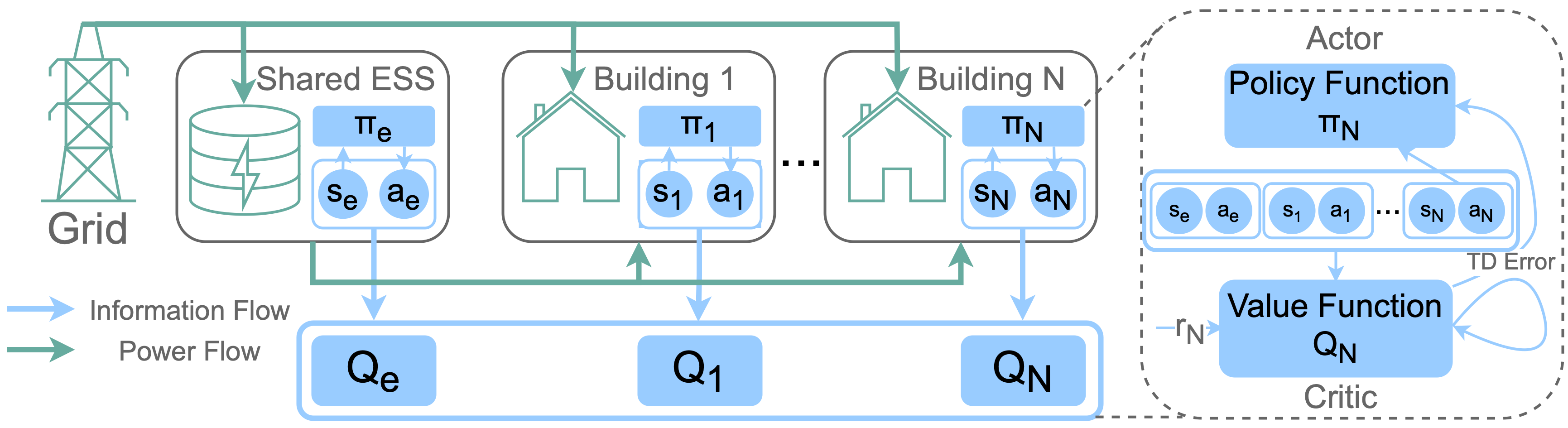}}
\vspace{-1.0em}
\caption{Schematic of proposed data-driven control strategies based on a multi-agent RL framework. A community-shared battery learns to cooperate with multiple buildings. Here the subscript $e$ denotes SESS, $1,\dots,N$ denotes buildings.}
\label{framwork}
\vspace{-1.0em}
\end{figure}

In our paper, we consider energy storage system which is set up to be controlled or dispatched by a utility or third party to maximize its service values for a given community. Similar to \cite{paridari2015demand}, we assume each building (customer) in a microgrid can schedule the shiftable loads based on user preferences and manage the interaction with the grid and the SESS. As HVAC system usually accounts for about half of building energy consumption \cite{wei2017deep}, and HVAC loads are flexible to provide incentives for users to use energy storage as the external power source during peak time, we choose to model HVAC as energy users in this paper. Fig.~\ref{framwork} demonstrates a building cluster equipped with SESS. The SESS charges from grid and dispatches its stored energy to buildings. Buildings schedule their power injection from both SESS and grid to minimize energy cost while keeping the indoor temperature within the preferred range of users.
%忘了画ess到building的连线，记得补上！

\vspace{-0.5em}
\subsection{HVAC Model}
\vspace{-0.5em}
Thermal resistor-capacitor networks (RC models) are a kind of typical models used to model the thermal dynamics of buildings \cite{wang2019development}. The indoor temperature of a building is influenced by heating transfer from HVAC systems and outdoor temperature. Assume that there is a set of $\mathcal{N}:=\{1,2,3, \ldots, N\}$ buildings and each of them has an HVAC system which operates in slotted time,  $\mathcal{K}:=\{1,2,3, \ldots, K\}$. Mathematically, RC model is represented as~\footnote{We note that the proposed RL method can be generalized to a variety of dynamics other than RC models.}
\begin{equation}
\vspace{-0.5em}
C_i \frac{d T_{i}^{in}}{d k}= \frac{T^{out}-T_{i}^{in}}{R_{i}}+w_{i}^{d} P_{i}^{d}+w_{i}^{g} P_{i}^{g};
\end{equation}
\begin{equation}
\vspace{-0.5em}
    P_{i,min}^{d}<P_{i}^{d}<P_{i,max}^{d};
\end{equation}
\begin{equation}
\vspace{-0.5em}
    P_{i,min}^{g}<P_{i}^{g}<P_{i,max}^{g};
\end{equation}
where $C_i$ is the thermal capacity of the $i^{th}$ building; $R_i$ is the thermal resistance between the $i^{th}$ building and outdoor; $T_i^{in}$ and $T^{out}$ represent the temperature of $i^{th}$ building and outdoor temperature respectively; $P_{i}^{d}$ and $P_{i}^{g}$ denote the heating input from SESS and grid along with weighting factors $w_{i}^{d}$ and $w_{i}^{g}$;  $P_{i,min}^{g}$, $P_{i,max}^{g}$, $P_{i,min}^{d}$ and $P_{i,max}^{d}$ are lower and upper power rate limitation. In our paper we consider both heating and cooling of the HVAC system, thus positive (negative) $P_{i}^{d}$ and $P_{i}^{g}$ means heating (cooling) input. In order to formulate the thermal dynamics to an Markov decision process (MDP) form, we give its discrete form:
    \begin{equation}
T_{i,k+1}^{in}= \delta_1 T_{i,k}^{in}+\delta_2 P_{i,k}^{d}+\delta_3 P_{i,k}^{g}+\delta_4 T_{k}^{out},
\end{equation}
where $\delta_{i}^{1}$, $\delta_{i}^{2}$, $\delta_{i}^{3}$, and $\delta_{i}^{4}$ are model parameters after discretization.

\vspace{-0.5em}
\subsection{SESS Model}
\vspace{-0.5em}
SESS schedules its charging and discharging from a real-time electricity market to minimize the charging cost while providing requested power to building customers. We assume that the SESS is a price taker which intends to charge at low prices and its operation will not affect the market prices \cite{wang2018energy}. SESS is modeled as follows:
\begin{equation}
\vspace{-1em}
    SOC_{k+1} = SOC_k + \delta_c c_k - \sum_{i}^{N}d_{i,k};
\end{equation}
\begin{equation}
\vspace{-0.5em}
     d_{i,k}=\frac{1}{\delta_d}\left|P_{i,k}^{d}\right|;
\end{equation}
\begin{equation}
\vspace{-0.5em}
    0\leq SOC_{k} \leq \bar{SOC};
\end{equation}
\begin{equation}
\vspace{-0.5em}
    0\leq d_{i,k} \leq \Bar{d_{i}};
\end{equation}
\begin{equation}
\vspace{-0.5em}
    0\leq c_{k} \leq \Bar{c};
\end{equation}
where $P_{i,k}^{d}$ denotes the power dispatch to the $i^{th}$ building at time k given SESS's discharging power $d_{i,k}$; $\Bar{SOC}$ denotes the capacity of SESS; $\Bar{c}$ and $\Bar{d_i}$ denotes the maximum charging/discharging power;  $c_k$ represents the charging power rate from grid at time $k$; $\delta_c$ and $\delta_d$ are charging and discharging efficiency term.

\vspace{-0.5em}
\subsection{Optimization problem}
\vspace{-0.5em}
The objectives of the optimization problem include minimizing the difference between building indoor temperature and users' preferred temperature as well as energy cost:

\begin{equation}
\vspace{-0.5em}
\begin{aligned}
\min _{\mathbf{c_k,P_{i,k}^{d},P_{i,k}^{g}}}
&\sum_{k \in \mathcal{K}} \sum_{i \in \mathcal{N}} \left( p_k \left(c_k+P_{i,k}^{g}\right)+\lambda \left|T_{i,k}^{in}-T_{i}^{target}\right| \right) \\
\text { s.t. } (2)-(9);
\end{aligned}
\end{equation}
where $p_k$ denotes the real-time price; $T_{i}^{target}$ represents the preferred temperature of the $i^{th}$ building user; and $\lambda$ is the weighting factor of optimization objectives.

However, due to the complicated dynamics involving temporally coupled SESS and multiple buildings, it is hard to solve the problem with conventional methods like dynamic programming. Moreover, as typical approaches like model predictive control (MPC) require accurate building and SESS models, while it is hard to model the systems exactly and implement distributed control in practice. Moreover, since the future price signal is always unknown, we need to design decision-makers which are informative of future price patterns. We provide an online MARL framework to solve the optimization problem mentioned above.

\vspace{-0.5em}
\section{Multi-Agent Learning Framework}
\vspace{-0.5em}
\subsection{MDP Process}
To solve the optimization problem with a MARL based approach, we firstly reformulate it as a multi-agent extension of MDP. 
%In this partially observable Markov game, there are totally $N+1$ agents for $N$ buildings and 1 SESS system. Here partially observable indicates that 
Each agent can only use its own state $\mathbf{s}_i$ during implementation, while we allow the use of full state $\mathbf{s}$ when training the RL policy. \\
%Thus the Markov game can be defined by a set of actions $\mathbf{a} = \mathbf{a}_1, \ldots, \mathbf{a}_{N+1}$, a set of states $\mathbf{s} = \mathbf{s}_1, \ldots, \mathbf{s}_{N+1}$ for each agent, a state transition function which produces the next state based on current state and actions of all agents $\mathcal{F}: \mathbf{s} \times \mathbf{a}_1 \times \ldots \times \mathbf{a}_{N+1} \mapsto \mathbf{s}$, and a reward function for each agent $r_i: \mathbf{s} \times \mathbf{a}_i \mapsto \mathbb{R}$. In this paper we use a policy gradient method to learn the policy of each agent directly, which maps the agent's own state to actions $\pi_{\mathbf{\theta}_i}: \mathbf{s}_i \mapsto \mathbf{a}_i$, where $\mathbf{\theta}$ represents the model parameters of policy (actor) network and will be omitted in the following statement. The aim of each agent is to maximize its own total expected reward $R_i=\sum_{k=0}^K \gamma^k r_i^k$, where $\gamma$ is a discount factor of the future reward. Next, we will give the state, action, reward function, and transition function regarding the optimization problem.\\
$\textbf{State}$: We use $\mathbf{s} = \mathbf{s}_1, \ldots, \mathbf{s}_{N+1}$ to denote a set of states for $N+1$ agents. For each building agent, $\mathbf{s}_{i,k} = \left(T_{i,k}^{in}, T_{k}^{out}, p_k, SOC_k\right)$. For SESS, $ \mathbf{s}_{SESS,k} = \left(T_{k}^{out}, p_k, \bar{p}_k,SOC_k\right)$. \\
$\textbf{Action}$: $\mathbf{a} = \mathbf{a}_1, \ldots, \mathbf{a}_{N+1}$ denotes a set of actions for $N+1$ agents. For each building agent, its actions include power input from the grid and discharged from SESS $\mathbf{a}_{i,k} = \left(P_{i,k}^{g}, P_{i,k}^{d}\right)$. For SESS, its action is the charging power $\mathbf{a}_{SESS,k} = \left(c_t \right)$. We use this design such that it is easier to train the SESS agent with simplified action space.\\
%$\textbf{Transtion Function}$: We build a simulator as the environment of reinforcement learning based on (2) - (9). With the input action and stored current state, the simulator returns next state and reward for each agent.\\
$\textbf{Reward Function}$: For each building agent, it aims to minimize the temperature difference between indoor temperature and user's preferred temperature and energy cost, thus we give negative reward based on the temperature difference and energy cost, respectively:
\begin{equation}
\vspace{-0.5em}
\label{building reward}
    r_{i, k}^{temp} = \left|T_{i,k}^{in}-T_{i}^{target}\right|; \quad    r_{i, k}^{energy} = p_k P_{i,k}^{g};
\end{equation}
\begin{equation}
\vspace{-0.5em}
    r_{i, k} = \alpha_{temp} r_{i, k}^{temp} + \alpha_{energy} r_{i, k}^{energy},
\end{equation}
where $\alpha_{temp}$ and $\alpha_{energy}$ are weighting factors of temperature difference and energy cost. For SESS agent, we encourage it to charge when the price is lower:
\begin{equation}
\vspace{-0.5em}
\label{SESS charging reward}
    r_{SESS,k} = \left(\bar{p}_k-p_k\right)c_k, k\in\left(1,K-1\right)
\end{equation}
where $\bar{p}_k$ is a weighted average of historical price: 
\begin{equation}
\vspace{-0.5em}
    \bar{p}_k=(1-\eta) \bar{p}_{k-1}+\eta p_k,
\end{equation}
where $\eta$ is a smoothing parameter to encode moving average \cite{wang2018energy}.
To prevent energy waste caused by unbalanced charging and discharging, we add a penalty for remaining energy in every testing episode:
\begin{equation}
\vspace{-0.5em}
\label{last step}
    r_{SESS,K} = \left(\bar{p}_k-p_k\right)c_k - \beta SOC_K,
\end{equation}
where $\beta$ is a negative penalty factor.

\vspace{-0.5em}
\subsection{MADDPG algorithm}
\vspace{-0.5em}
The main idea of Multi-Agent Deep Deterministic Policy Gradient (MADDPG) algorithm is adopting \emph{centralized training with decentralized execution}. Each agent has both actor and critic neural network. Given the states and actions of all agents defined in the previous subsection, the critic network is a value function to estimate
expected discounted returns. It is demonstrated in \cite{lowe2017multi} that given the actions of other agents when training critic network, the environment is stationary even as the policies change: $P_r\left(s^{\prime} \mid s, a_1, \ldots, a_N, \pi_1, \ldots, \pi_N\right)=P_r\left(s^{\prime} \mid s, a_1, \ldots, a_N\right)=P_r\left(s^{\prime} \mid s, a_1, \ldots, a_N, \pi_1^{\prime}, \ldots, \pi_N^{\prime}\right)$, where $P_r$ denotes the state transition probability. With a well-trained value function, the actor network, which is utilized to map states to actions, updates its parameter by policy gradient method under the critic's guidance.

At every time step $k$, each agent takes actions based on its own states. Without loss of generality, we use $i$ as the index of building agents and SESS agent:
$    \mathbf{a}_{i,k}=\pi_{\theta_i}\left(\mathbf{s}_{i,k}\right)$, where $\pi_{\theta_i}$ represents the actor network's policy with model parameters $\theta_i$, and $\theta_i$ will be omitted in the following statement. Given the actions of all agents and their current state, the environment will feedback reward and next state for all agents. Thus we can obtain a tuple $\left(\mathbf{s},\mathbf{a},\mathbf{s}^{\prime},\mathbf{r}\right)$ and store it in the replay buffer $\mathcal{D}$, where $\mathbf{s}^{\prime}$ denotes the full state in next time step. (Here we omit the time index without loss of generality.)

With the tuples sampled from experience replay buffer, the value function is updated by minimizing the following loss:
\begin{equation}
\label{critic loss}
\mathcal{L}\left(\mu_i\right)=\mathbb{E}_{\mathbf{s}, \mathbf{a}, r, \mathbf{s}^{\prime}}\left[\left(Q_i\left(\mathbf{s}, \mathbf{a}\right)-y\right)^2\right],
\end{equation}
where $y=r_i+\left.\gamma Q_i^{\prime}\left(\mathbf{s}^{\prime}, \mathbf{a}^{\prime}\right)\right|_{\mathbf{a}_j^{\prime}=\mathbf{\pi}_j^{\prime}\left(\mathbf{s}_j\right)}$; $\mu$ is the model parameters of critic network; $Q_i$ and $Q_i^{\prime}$ represent critic network and target critic network. %\footnote{We use a soft update method in the actor-critic framework, thus each actor and critic network has corresponding target network \cite{lowe2017multi}.} w.r.t. model parameters $\mu$ and $\mu^{\prime}$, respectively, 
$j$ is the index of mini-batch samples sampled from $\mathcal{D}$ in each episode, which will be further introduced in Algorithm \ref{A1}.

The policy gradient method used to update actor network is given as follows:
\begin{equation}
\label{actor update}
\nabla_{\theta_i} J \left( \pi_i \right)=E_{\mathbf{s}, \mathbf{a} \sim \mathcal{D}}\left[\left.\nabla_{\theta_i} \pi_i\left(\mathbf{a}_i \mid \mathbf{s}_i\right) \nabla_{\mathbf{a}_i} Q_i\left(\mathbf{s}, \mathbf{a}\right)\right|_{\mathbf{a}_i=\pi_i\left(\mathbf{s}_i\right)}\right].
\end{equation}

\begin{algorithm}[h]
    \caption{Learning to schedule coupled SESS and HVAC}
    \label{A1}
    \begin{algorithmic}[1]
    \REQUIRE The number of buildings $N$;  $p_k$,  $T_{k}^{out}, k\in\left[1,K\right]$
    \STATE Initialize experience replay buffer $\mathcal{D}$ and environment;
    %\STATE Initialize the environment for $N+1$ agents;
    \FOR{episode =  $1,2,\dots,M$}
    \STATE Reset environment;
    \FOR{step =  $1,2, \dots, K$}
    \STATE Each agent select actions $\mathbf{a}_{i, k}=\pi_{\theta_i}\left(\mathbf{s}_{i, k}\right)$
    \STATE Execute $\mathbf{a}_{i,k}$; observe $\mathbf{s}_{i,k+1}$ and reward $r_{i,k+1}$;
    \STATE Store tuple $\left( \mathbf{s_k}, \mathbf{a_k}, \mathbf{s_{k+1}}, \mathbf{r_k} \right)$ in $\mathcal{D}$;
    \STATE Update state: $\mathbf{s_k} \leftarrow \mathbf{s_{k+1}}$
    \FOR{agent = $1,2,\dots,N+1$}
    \STATE Sample mini-batch $\mathbf{B}$ with $B_{size}$ tuples $\left( \mathbf{s_k}, \mathbf{a_k}, \mathbf{s_{k+1}}, \mathbf{r_k} \right)$ from $\mathcal{D}$;
    \STATE Calculate actions with target actor network: $\mathbf{a}_j^{\prime}=\boldsymbol{\pi}_j^{\prime}\left(\mathbf{s}_j\right)$ for $j$ in mini-batch $\mathbf{B}$;
    \STATE Calculate $Q_{i}^{\prime}\left(\mathbf{s}_j,\mathbf{a}_j\right)$ for $j$ in mini-batch $\mathcal{B}$;
    \STATE Update critic network by minimizing \eqref{critic loss};
    \STATE Calculate $Q_{i}\left(\mathbf{s}_j,\mathbf{a}_j\right)$ for $j$;
    \STATE Update actor network with \eqref{actor update}; Update target network%: $\theta_i^{\prime} \leftarrow \tau \theta_i+(1-\tau) \theta_i^{\prime}$, $\mu_i^{\prime} \leftarrow \tau \mu_i+(1-\tau) \mu_i^{\prime}$;
    \ENDFOR
    \ENDFOR
    \ENDFOR

    \end{algorithmic}
    \end{algorithm}

\vspace{-0.5em}
\section{Case Study}
\vspace{-0.5em}
\subsection{Experiment Setup}
\vspace{-0.5em}
The real-world weather and pricing dataset is from Hourly energy demand generation and weather$\footnote{https://www.kaggle.com/datasets/nicholasjhana/energy-consumption-generation-prices-and-weather?select=energy\_dataset.csv}$, a dataset in Kaggle. It contains Spain's 4 years of electrical consumption, generation, pricing, and weather data. Each episode contains 96 time steps (i.e. 4 days) and the time interval is 1 hour. To test the framework's performance in different situations, we show 3 typical cases here: (1) Winter (case 1): the HVAC systems need to heat all the time; (2) Spring (case 2): the HVAC system should both heat and cool; (3) Summer (case 3): the HVAC system needs to cool to keep comfort temperature. Other parameters we use are given in Table \ref{parameter table}. In this section, we show the results of 3 cases in Table \ref{results} and visualize some typical results of case 2 with $\alpha_{temp}/ \alpha_{energy}=10$ in \texttt{Proposed} of the most interest.

\begin{table}
\vspace{-1.0em}
\caption{Parameters used for SESS, HVAC, and RL simulations.}
\centering
\begin{tabular}{ll|ll|ll} 
\hline
$T_{target}\left(^{\circ}C\right)$ & 20  & $w_2^g$                    & 1   & $\lambda$                   & 0.3   \\
$R_1\left(^{\circ}C/kW\right)$     & 8   & $P_1^{max}\left(kW\right)$ & 5   & $\Bar{SOC}\left(kWh\right)$ & 10    \\
$R_2\left(^{\circ}C/kW\right)$     & 6   & $P_1^{min}\left(kW\right)$ & -5  & $\Bar{d_i}\left(kW\right)$  & 5     \\
$C_1\left(kWh/^{\circ}C\right)$    & 15  & $P_2^{max}\left(kW\right)$ & 5   & $\Bar{c}\left(kW\right)$    & 5     \\
$C_2\left(kWh/^{\circ}C\right)$    & 14  & $P_2^{min}\left(kW\right)$ & -5  & $\gamma$                    & 0.9   \\
$w_1^d$                            & 0.9 & $\delta_c$                 & 0.9 & $\tau$                      & 0.01  \\
$w_1^g$                            & 1.1 & $\delta_d$                 & 1.1 & $N$                         & 2     \\
$w_2^d$                            & 1   & $\eta$                     & 0.2 & $Episode$                   & 500   \\
\hline
\end{tabular}
\label{parameter table}
\vspace{-1.0em}
\end{table}

\begin{figure}[htbp]
\vspace{-1.3em}
\centerline{\includegraphics[width=0.45\textwidth]{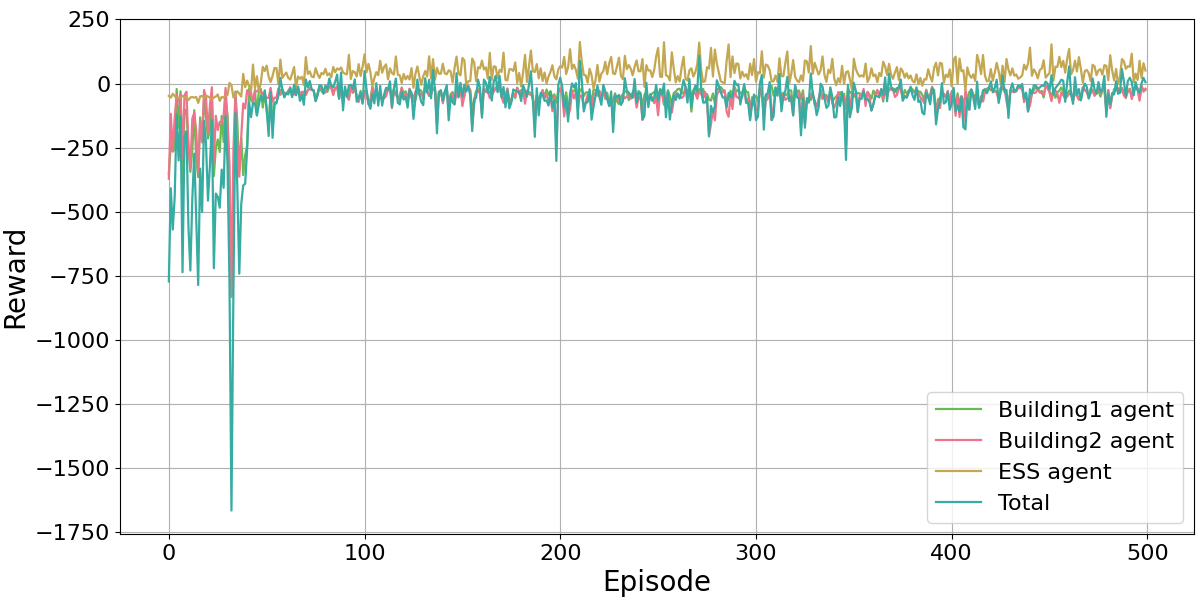}}
\vspace{-1.0em}
\caption{Training reward evolution of \texttt{Proposed} in case 2. }
\label{reward}
\vspace{-1.0em}
\end{figure}

\begin{table*}
\vspace{-1.0em}
\caption{Comparison of experiment results.}
\centering
\setlength{\tabcolsep}{4pt}
\begin{tabular}{l|lllllllll} 
\hline
\multicolumn{1}{c|}{\multirow{2}{*}{Methods}} & \multicolumn{3}{c}{Case1}                         & \multicolumn{3}{c}{Case2 }                        & \multicolumn{3}{c}{Case3}                          \\
\multicolumn{1}{c|}{}                   & \multicolumn{1}{c}{ATD} & \multicolumn{1}{c}{TEC} & \multicolumn{1}{c}{CP}  & \multicolumn{1}{c}{ATD} & \multicolumn{1}{c}{TEC} & \multicolumn{1}{c}{CP} & \multicolumn{1}{c}{ATD} & \multicolumn{1}{c}{TEC}  & \multicolumn{1}{c}{CP}\\ 
\ChangeRT{0.6pt}
\texttt{Heuristics}                                      & 0.113          & 21.260          & 0.613          & 0.109          & 19.793         &0.605          & \textbf{0.107} & 22.200          &0.558                  \\
\texttt{User Only}                                       & 0.491          & 14.574          & 0.833          & 0.098          & 11.791         &0.392          & 0.181          & 14.280          &0.420                  \\
\texttt{Centralized}                                     & 0.501          & 20.277          & 0.977          & 0.520          & 15.531         &0.892          & 0.917          & 22.045          &0.997                  \\
\texttt{Proposed($\alpha_{temp}$/ $\alpha_{energy}$=10)} & 0.256          & \textbf{14.237} & 0.590          & 0.117          & \textbf{9.247} &0.346          & 0.377          & \textbf{12.698} &0.492                  \\
\texttt{Proposed($\alpha_{temp}$/ $\alpha_{energy}$=20)} & 0.184          & 16.576          & 0.573          & \textbf{0.076} & 11.438         &0.362          & 0.189          & 13.985          &0.418                  \\
\texttt{Proposed($\alpha_{temp}$/ $\alpha_{energy}$=25)} & \textbf{0.088} & 17.746          & \textbf{0.505} & 0.081          & 9.752          &\textbf{0.324} & 0.147          & 14.645          &\textbf{0.410}         \\

\ChangeRT{0.6pt}
\end{tabular}
\label{results}
\end{table*}

\begin{figure}[htbp]
\vspace{-1.0em}
\centerline{\includegraphics[width=0.45\textwidth]{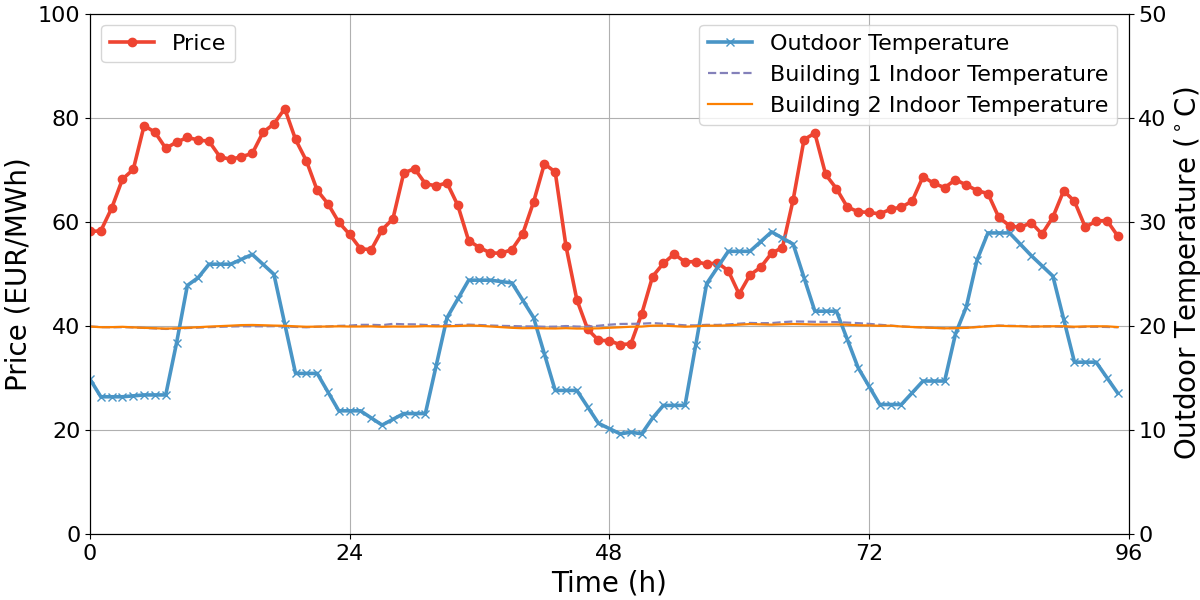}}
\vspace{-1.0em}
\caption{The external inputs of price and outdoor temperature, along with the indoor temperature under \texttt{Proposed}.}
\label{external}
%\vspace{-1.0em}
\end{figure}

\vspace{-0.5em}
\begin{figure}[htbp]
\vspace{-0.5em}
\centerline{\includegraphics[width=0.45\textwidth]{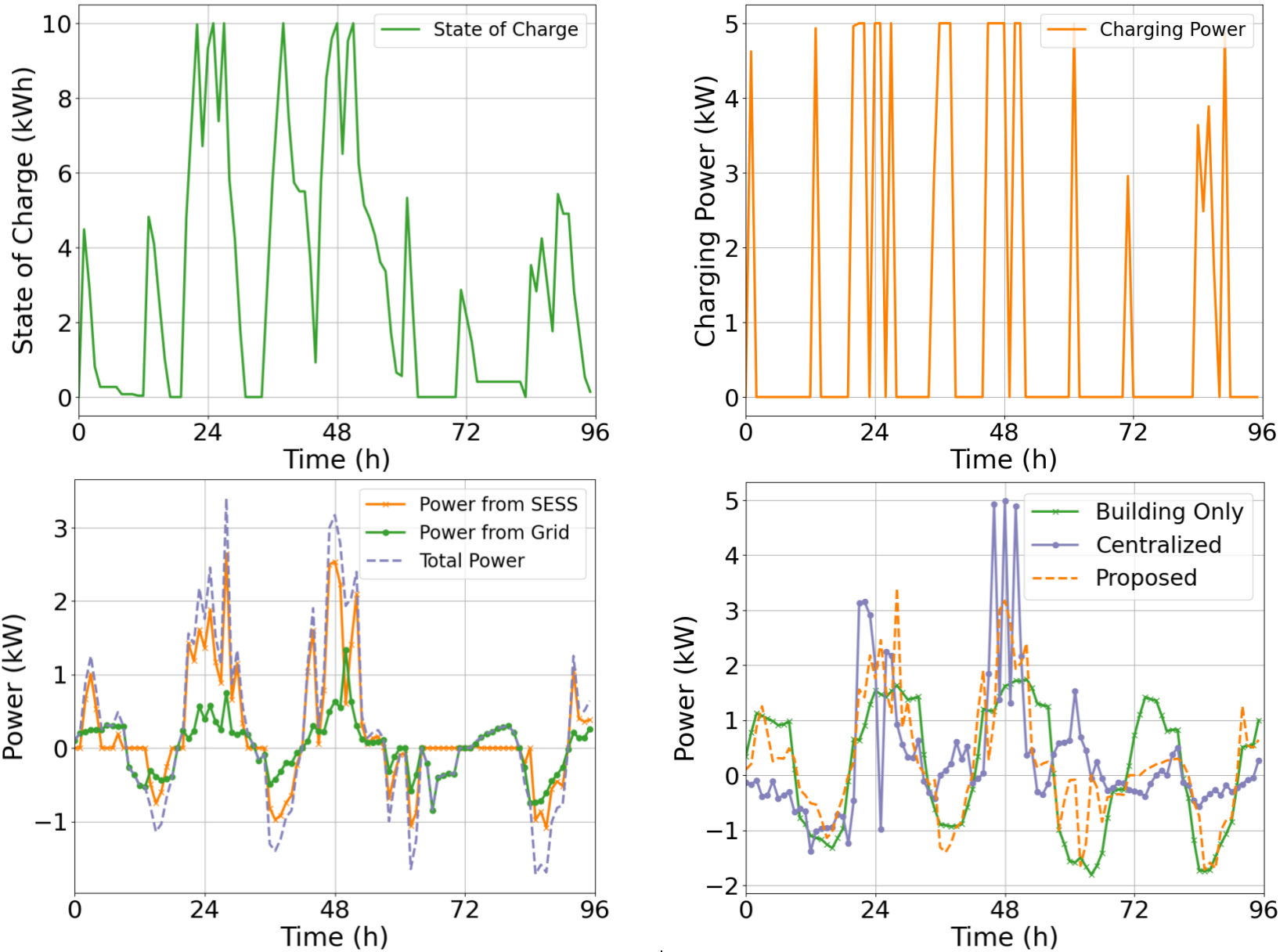}}
\vspace{-1.0em}
\caption{Simulation results in case 2. (a) State-of-charge of SESS under \texttt{Proposed}; (b) charging power $c_k$ of SESS under \texttt{Proposed}; (c) power injection of building 1 from grid $P_{1,k}^{g}$, SESS $P_{1,k}^{d}$, and total power injection $P_{1,k}^{g}+P_{1,k}^{d}$ under \texttt{Proposed}; (d) total power injection of building 1 under \texttt{Building Only}, \texttt{Centralized}, and \texttt{Proposed}.}
\label{test marl}
\vspace{-1.0em}
\end{figure}

\begin{figure}[htbp]
\vspace{-0.5em}
\centerline{\includegraphics[width=0.45\textwidth]{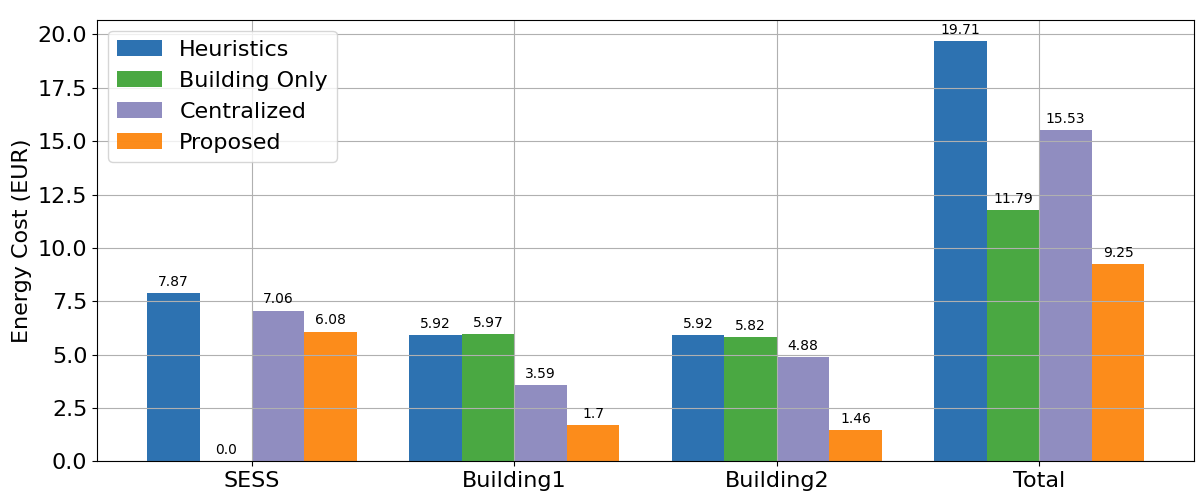}}
\vspace{-1.0em}
\caption{Energy Cost comparison.}
\label{energy cost}
\vspace{-1.0em}
\end{figure}

Here we choose 3 baselines to compare with our proposed method ($\texttt{Proposed}$):\\
\textbf{Heuristic Scheme} (\texttt{Heuristics}): \texttt{Heuristics} uses some rules based on SOC, price, and outdoor temperature to control the power rate. The charging power of SESS $c_k = 5 kW$ if $p_k<\Bar{p}_k$, otherwise $c_k = 0$. The power injection of HVAC systems from grid and SESS $P_{i,k}^d=1$ and $P_{i,k}^g=1$ if outdoor temperature $T_{k}^{out}<0$, otherwise $P_{i,k}^d=-1$ and $P_{i,k}^g=-1$. The charging rate should not obey the constraint of SESS's capacity (7).\\
\textbf{HVAC system operates without SESS} (\texttt{User Only}): This scheme only considers independent HVAC operators. In other words, we trained building agents separately.\\ 
\textbf{Centralized Reinforcement learning} (\texttt{Centralized}): \texttt{Centralized} trains a single agent with combined states for SESS and HVAC systems. %To be specific, the state of this reinforcement learning agent contains $\left(T_{1,t}^{in},\dots,T_{N,t}^{in},T_{t}^{out},p_t,\Bar{p_t},SOC_t\right)$, while action contains $\left(c_t,P_{1,t}^d,\dots,P_{N,t}^d,P_{1,t}^g,\dots,P_{N,t}^g\right)$.

\vspace{-0.5em}
\subsection{Performance Analysis}
\vspace{-0.5em}
We introduce 2 performance metrics to analyze the performance numerically \cite{yu2020multi}: the \textbf{Average Temperature Deviation (ATD)} $ATD=\frac{1}{KN}\sum_{k=1}^{K}\sum_{i=1}^{N}\left|T_{i,k}-T_{target}\right|$; \textbf{Total Energy Cost (TEC)} $TEC=\sum_{k=1}^{K}\sum_{i=1}^{N}\left(\left|P_{i,k}^{g}\right|+d_{i,k}\right)$; and \textbf{Comprehensive Performance (CP)} $CP = 0.5 \frac{ATD}{ATD_{max}}+ 0.5 \frac{TEC}{TEC_{max}},$ where $ATD_{max}$ and $TEC_{max}$ are the maximum ATD and TEC among each case, respectively.

In TABLE \ref{results}, we present the simulation results of all cases with all baseline algorithms and \texttt{Proposed}. The TEC of \texttt{Proposed} is the smallest in all 3 cases; although the ATD is not the smallest, it is smaller than $0.4^{\circ}C$, which is a relatively small value in the real world. The performance on metric CE indicates that \texttt{Proposed} with user-tunable weighting parameter $\alpha_{temp}$ and $\alpha_{energy}$ can achieve the heterogeneous users' control objectives and always has the potential to obtain the best comprehensive temperature control and energy cost saving performance once we carefully set the $\alpha_{temp}$ and $\alpha_{energy}$. %Next, we will first analyze how our proposed algorithm achieves the best performance on energy cost saving and at the same keeps a relatively small temperature deviation. Then we will further compare with baseline methods to show our proposed algorithm's advantages in energy cost saving in the next section.

Fig. \ref{reward} suggests that the reward of all 3 agents converge after around 50 episodes with \texttt{Proposed}. Recall in Equation \eqref{SESS charging reward}, the SESS is encouraged to charge when real-time price $p_k$ is lower than weighted average price $\Bar{p}_k$ and keeps $c_k=0$ to avoid the negative reward (penalty) when $p_k<\Bar{p}_k$. It can be observed in Fig. \ref{test marl}(b) that the SESS actively charges to take advantage of the price signal shown in Fig. \ref{external} with the encouragement of our well-designed reward. In our proposed method, the 3 agents tend to maximize their own reward to achieve the greater total reward. In other words, the real-time power injection of Fig. \ref{test marl}(c) shows that the building agent determines the heating or cooling power according to the temperature situation and determines the source of energy based on the real-time SOC and price to take advantage of the SESS. For instance, from Fig. \ref{external} and Fig. \ref{test marl} we can see that at the $24^{th}$ hour and the $48^{th}$ hour, due to the lower temperature, the HVAC needs a larger heating power to maintain the temperature to reduce the negative reward in \eqref{building reward} regarding the temperature difference. 

We observe that our well-designed settings ensure the balance of battery charge and discharge as shown in Fig. \ref{test marl}(a), and use the energy storage of the battery as much as possible to minimize energy costs. One of the settings is that the discharge (corresponding to $P_{i,k}^{d}$) of the SESS is controlled by the building agent, which can not only avoid complex communication problems but also enable the building to perform power dispatch according to its real-time energy demand as illustrated in Fig. \ref{test marl}(c). Another unique design is to avoid excessive energy storage of the SESS by using Equation \eqref{last step}.

As shown in Table \ref{results}, \texttt{Heuristics} has an energy cost about 2 times that of \texttt{Proposed} in case 2. This is because the \texttt{Heuristics} is only based on the current state, and it cannot learn some complex strategies.
%, such as maintaining the balance of charge and discharge of SESS to avoid energy waste caused by excess energy storage, and continuous power values to avoid frequent heating or cooling. 
As for \texttt{User Only}, from Fig. \ref{test marl}(d), its power curve is similar to that of \texttt{Proposed}. However, the energy consumption in each case is still about $2\%$ to $28\%$ higher than that of \texttt{Proposed}. Although the rewards of RL in \texttt{User Only} can converge well, due to the lack of SESS, %when the price is high and the electricity demand is also high, the agent can only make a trade-off between energy saving and guaranteed performance, and there is no way to use the energy storage of SESS. 
as illustrated in Fig. \ref{energy cost}, the energy cost of building 1 and building 2 of \texttt{Proposed} is much lower than that of \texttt{User Only}. Speaking of \texttt{Centralized}, it can be seen from Table \ref{results} that it is not as good at maintaining temperature and saving energy as \texttt{Proposed}. This is because the convergence of RL is difficult due to the increased dimension of the state and action spaces. In addition, due to the existence of multiple rewards in centralized training, it is difficult for the agent to learn to trade off among multiple rewards to maximize return.

%\begin{figure*}[htbp]
%  \centering
%  \subfigure[SOC]{
%    \label{soc} 
%    \includegraphics[scale=0.25]{SOC.png}}
%  \hspace{-0.3in}
%  \subfigure[Charging Power]{
%    \label{charging} 
%    \includegraphics[scale=0.25]{marl_charging.png}}
%  \hspace{-0.3in}
%  \subfigure[Power Injection of Building 1]{
%    \label{Building1} 
%    \includegraphics[scale=0.25]{Building1_result.png}}
%  \hspace{-0.3in}
%  \subfigure[Total Power Injection]{
%    \label{Building2} 
%    \includegraphics[scale=0.25]{all_power.png}}
%  \caption{test marl}
%  \label{marlmarl} 
%\end{figure*}

\vspace{-0.5em}
\section{Conclusion and Future Works}
In this paper, we design a multi-agent, distributed power dispatch and information exchange framework for SESS and a group of users in the community. With the multi-agent reinforcement learning algorithm, each SESS and user agent can decide the power dispatch based on its own states to meet the energy demand of HVAC system and SESS’s charging/discharging balance without additional communication. The well-designed reward functions help the agents to decide energy demand based on temperature deviation and take advantage of the capacity of SESS and price signals to save energy costs. Numerical experiments with real-world weather and electricity prices data show that the agents achieve great energy cost savings and energy demand satisfaction in all simulation cases. We will consider a more scalable framework and algorithm in our future work.

\vspace{-0.5em}
\bibliographystyle{IEEEtran}
\bibliography{bib}
\end{document}